\documentclass{pasj00}

\begin{document}
\SetRunningHead{Yan et al.}{The estimate of emission region
locations of {\it Fermi} FSRQs}

\title{The estimate of emission region locations of {\it Fermi} flat spectrum radio quasars }

\author{Dahai \textsc{Yan}, Houdun \textsc{Zeng}, and Li \textsc{Zhang} }
\affil{Department of physics,Yunnan University,
    Kunming, China}
 \email{lizhang@ynu.edu.cn}

\KeyWords{galaxies: active -- radiation mechanisms: nonthermal --  quasars: general -- $\gamma$-rays: observations}

\maketitle

\begin{abstract}
We study the locations of emission regions through modelling the
quasi-simultaneous multi-frequency spectral energy distributions
of 21 {\it Fermi} flat spectrum radio quasars (FSRQs) in the frame
of a multi-component one-zone leptonic model. In our calculations,
we take the detailed broad line region (BLR) structure into
account and discuss the effect of the uncertainty of the BLR
structure on constraining the location of the emission regions for
each FSQR, meanwhile we also include both the internal and
external absorptions. Our results indicate that (1) the
contribution of external Compton-BLR component is important to
$\gamma$-ray emission, and the energy density of external target
photon fields depends on the location of the emission region,
which can be derived through reproducing the observed $\gamma$-ray
emission, and (2) the emission regions of FSRQs with relative low
accretion disk luminosity lie in the region of $7.9\times10^{16}$\
--\ $1.3\times10^{18}$\ cm (300 -- 4300 Schwarzschild radii) from
central black hole, and for FSRQs with high accretion disk
luminosity, the emission regions locate in the larger region of
$2.6\times10^{17}$\ --\ $4.2\times10^{18}$\ cm (300 -- 5600
Schwarzschild radii).
\end{abstract}

\section{Introduction}

Blazars are the most extreme class of active galactic nuclei
(AGNs). Their spectral energy distributions (SEDs) are
characterized by two distinct bumps: the first bump located at
low-energy band is dominated by the synchrotron emission of
relativistic electrons which originates in a relativistic jet, and
the second bump located at high-energy band could be produced by
inverse Compton (IC) scattering (e.g., B\"{o}ttcher 2007). Various
soft photon sources seed the synchrotron self-Compton (SSC)
process (e.g., Rees 1967; Maraschi et al. 1992) and external
Compton (EC) process (e.g., Dermer \& Schlickeiser 1993; Sikora et
al. 1994) in the jet to produce $\gamma$-rays. It should be noted
that hadronic models have also been proposed to explain the
multi-band emissions of blazars (e.g., Mannheim 1993; M\"{u}cke et
al. 2003). At present, the detectors including the ground-based
and satellite-based ones can provide us the high quality data to
construct SEDs. Therefore, the simultaneous and quasi-simultaneous
SEDs become the basic tools to study the physics properties of
blazars, such as magnetic field, the size of emission region,
Doppler factor and so on.

The Large Area Telescope (LAT) onboard the {\it Fermi} satellite
provides unprecedented sensitivity in the $\gamma$-ray band. In
its two years operations, 987 blazars were detected, including 360
flat spectrum radio quasars (FSRQs), 423 BL Lacs and 204 unknown
type blazars \citep{Ackerman11}. With {\it Fermi}, {\it Swift} and
other observatories, \citet{abdo10a} assembled high quality data
of 48 blazars in the first three months LAT sample (LAT Bright AGN
Sample: LBAS) to build quasi-simultaneous SEDs. The data from {\it
Swift} were collected in one day or several days, however, the
{\it Fermi}-LAT data have been averaged over a three months
period. Therefore, the multi-frequency data are quasi-simultaneous
but not really simultaneous. Very recently, \citet{Giommi11} built
the SEDs for 105 blazars with the simultaneous {\it Planck}, {\it
Swift}, {\it Fermi}, and ground-based data. The data of the 105
blazars are obtained between 2009 Dec. and 2010 Oct., which were
all collected in one day or several days, and the multi-frequency
data are really simultaneous.

Since the {\it Fermi} data are publicly available, many authors
have studied the physical properties of blazars with
multi-frequency data from radio-X-ray to $\gamma$-ray bands.
\citet{ghisellini10} constructed the SEDs for 85 blazars in LBAS,
and found that there is a positive correlation between the jet
power and the luminosity of the accretion disk in FSRQs. Later,
\citet{ghisellini11} also analyzed the physical properties of
blazars with high redshifts in the 11 months {\it Fermi} AGN
sample (the First LAT AGN Catalog: 1LAC) by using the
multi-frequency data. However, most of the multi-frequency data
used by these authors are not simultaneous or quasi-simultaneous.

In the study of $\gamma$-ray emissions of blazars, one of
important physical properties is the locations of the $\gamma$-ray
emission regions. Many authors studied the locations of emission
regions in different ways. For example, observed time lag of the
$\gamma$-ray emission relative to broad emission lines for 3C 273
is used to constrain the $\gamma$-ray emission region
\citet{Liu11}, the $\gamma$-ray absorptions by internal photons of
AGNs can be used to constrain the locations of the $\gamma$-ray
emission regions \citep{sia,bai,pout}, and the $\gamma$-ray flux
variability of a blazar can infer its location of $\gamma$-ray
emission \citep{tavecc10}. In this paper, we study the locations
of the emission regions of 21 blazars whose simultaneous and
quasi-simultaneous multi-frequency SEDs have been given by
\citet{Giommi11} and \citet{abdo10a} in the frame of the
multi-component model described in \citet{dermer09}.

In this study, we use cosmological parameters ($H_0, \Omega_m,
\Omega_{\Lambda}$) = (70 km s$^{-1}$ Mpc$^{-1}$, 0.3, 0.7).

\section{Photon Emission and Absorption of FSRQs}

\subsection{Photon Emission}

Broad band emission from blazars can be usually described in the
framework of leptonic models. Since there is difficulty to explain
SEDs of FRSQs in a simple one-zone SSC model \citep{abdo10a}, we
study the photon emission properties of FSRQs in the frame of
one-zone, homogeneous synchrotron and inverse Compton model given
by \citet{dermer09}. There are three soft photon seeds in this
model: synchrotron photons, accretion disk photons, and broad line
region (BLR) photons.

Nonthermal photon emission is assumed to be produced by both the
synchrotron radiation and IC scattering of relativistic electrons
in a spherical blob of the jet which is moving relativistically at
a small angle to our line of sight, and the observed radiation is
strongly boosted by a relativistic Doppler factor $\delta_D$. The relativistic electron distribution which we
used here is the same as that given by \citet{dermer09}, i.e.
\begin{eqnarray}
N_{\rm e}^{\prime}(\gamma^{\prime})=K_{\rm
e}^{\prime}H(\gamma^{\prime};\gamma_{\rm min}^{\prime},\gamma_{\rm
max}^{\prime})\{{\gamma^{\prime
-p_1}\exp(-\gamma^{\prime}/\gamma_{\rm b}^{\prime})}
\nonumber \\
\times H[(p_{\rm 2}-p_{\rm 1})\gamma_{\rm
b}^{\prime}-\gamma^{\prime}]+[(p_{\rm 2}-p_{\rm 1})\gamma_{\rm
b}^{\prime}]^{p_{\rm 2}-p_{\rm 1}}\gamma^{\prime -p_{\rm 2}}
\nonumber \\
\times \exp(p_{\rm 1}-p_{\rm 2})H[\gamma^{\prime}-(p_{\rm
2}-p_{\rm 1})\gamma_{\rm b}^{\prime}]\},
\end{eqnarray}
where $K_{\rm e}^{\prime}$ is the normalization factor,
$H(x;x_{1},x_{2})$ is the Heaviside function: $H(x;x_{1},x_{2})=1$
for $x_{1}\leq x\leq x_{2}$ and $H(x;x_{1},x_{2})=0$ everywhere
else; as well as $H(x)=0$ for $x<0$ and $H(x)=1$ for $x\geq0$. In
the co-moving frame, this distribution is a double power law with
two energy cutoffs: $\gamma_{\rm min}^{\prime}$ and $\gamma_{\rm
max}^{\prime}$. The spectrum is smoothly connected with indices $
p_1$ and $p_2$ below and above the electrons' break energy
$\gamma_{\rm b}^{\prime}$. Note that here and throughout the
paper, unprimed quantities refer to the distant observer's frame
on Earth and primed ones refer to the co-moving frame.

After giving the electron distribution, the synchrotron flux can
be given by Finke et al. (2008)
\begin{equation}
\nu F^{\rm syn}_{\nu}=\frac{\sqrt{3}\delta^4_{\rm
D}\epsilon'e^3B}{4\pi h d^2_{\rm
L}}\int^\infty_0d\gamma'N'_e(\gamma')R(x)\;\;,
\label{syn}
\end{equation}
where $e$ is the electron charge, $B$ is the magnetic field
strength, $h$ is the Planck constant, and $d_{\rm L}$ is the
distance to the source with a redshift $z$. Here $m_{\rm
e}c^2\epsilon^{\prime}=h\nu(1+z)/\delta_{\rm D}$ is synchrotron
photons energy in the co-moving frame, where $m_{\rm e}$ is the
rest mass of electron and $c$ is the speed of light. In equation
(\ref{syn}),
$R(x)=(x/2)\int^\pi_0d\theta\sin\theta\int^\infty_{x/\sin\theta}dt
K_{5/3}dt$, where $x=4\pi\epsilon'm^2_ec^3/3eBh\gamma^{'2}$,
$\theta$ is the angle between magnetic field and velocity of high
energy electrons, and $K_{5/3}(t)$ is the modified Bessel function
of the second kind of order 5/3. Here we use an approximation for
$R(x)$ given by Finke et al. (2008). The synchrotron photons with
$\nu F^{\rm syn}_\nu$ have following spectral energy density
\begin{equation}
u^{\prime}_{\rm syn}(\epsilon^{\prime})=\frac{3d_{\rm L}^2\nu
F_{\nu}^{\rm syn}}{cR_{\rm b}^{\prime 2}\delta_{\rm
D}^4\epsilon^{\prime}}\;,
\end{equation}
where $R_{\rm b}^{\prime}=\frac{c\delta_{\rm D}t_{\rm v,min}}{1+z}$ is
blob's radius and $t_{\rm v,min}$ is minimum variability timescale.

For the IC scattering, there are two kinds of soft photon fields:
non-thermal and thermal. The non-thermal field is the synchrotron
photons field and high energy peak is created in the SSC
process. For isotropic and homogeneous photon and electron
distributions, the SSC flux, $\nu F_{\nu}$, is given by Finke et
al. (2008)
\begin{eqnarray}
\nu F^{\rm SSC}_{\nu}&=&\frac{3}{4}c\sigma_{\rm
T}\epsilon_{s}^{\prime 2}\frac{\delta_{\rm D}^4}{4\pi d_{\rm
L}^2}\int_0^{\infty} d\epsilon^{\prime}\frac{u^{\prime}_{\rm
syn}(\epsilon^{\prime})}{\epsilon^{\prime 2}}\\\nonumber
&&\int_{\gamma_{\rm min}^{\prime}}^{\gamma_{\rm max}^{\prime}}
d\gamma^{\prime}\;
\frac{N^{\prime}_e(\gamma^{\prime})}{\gamma^{\prime 2}}F_{\rm
C}(q^{\prime},\Gamma_{\rm e}^{\prime})\;,
\end{eqnarray}
where $\sigma_{\rm T}$ is the Thomson cross section, $m_{\rm
e}c^2\epsilon^{\prime}_{s}=h\nu(1+z)/\delta_{\rm D}$ is the energy
of IC scattered photons in the co-moving frame, $F_{\rm
C}(q^{\prime},\Gamma_{\rm e}^{\prime})=2q^{\prime}{\rm
ln}q^{\prime}+(1+2q^{\prime})(1-q^{\prime})+\frac{q^{\prime
2}\Gamma_{\rm e}^{\prime 2}}{2(1+q^{\prime}\Gamma_{\rm
e}^{\prime})}(1-q^{\prime})$, $
q^{\prime}=\frac{\epsilon^{\prime}/\gamma^{\prime}}{\Gamma^{\prime}_{\rm
e}(1-\epsilon^{\prime}/\gamma^{\prime})}$, $\Gamma_{\rm
e}^{\prime}=4\epsilon^{\prime}\gamma^{\prime}$, and $
\frac{1}{4\gamma^{\prime 2}}\leq q\leq1$.

The thermal soft photon field includes soft photon direct from
accretion disk and broad line region (BLR) photon components and
the IC process is called external Compton (EC) process. Here, we
do not take the BLR line radiation into account but only consider
the BLR Thomson scattered photon field. Assuming a standard
accretion disk with a central black hole mass $M=M_{8}10^8M_{\rm
sun}$ and an accretion disk luminosity $L_{\rm d}$, its mean
photon energy in units of $m_ec^2$ at radius $R$ is $
\epsilon_{\rm d}(R)=1.5\times10^{-4}(\frac{10\ell_{\rm Edd}} {M_8\
\eta\ \varphi(R)})^{\frac{1}{4}}(\frac{R}{r_{\rm
g}})^{-\frac{3}{4}}$ \citep{finke102}, where $\ell_{\rm
Edd}=\frac{L_{\rm d}}{L_{\rm Edd}}$ with $L_{\rm
Edd}=1.26\times10^{46}M_8 \rm \,erg\ s^{-1}$, $\eta$ is the
accretion efficiency, $r_{\rm
g}=\frac{GM}{c^2}\cong1.5\times10^{13}M_8 \rm \,cm$, $
\varphi(R)=1-\beta_{\rm i}(\frac{R_{\rm in}^{\rm d}}{R})^{0.5}$,
$\beta_{\rm i}\cong1.0$, and $R_{\rm in}^{\rm d}=6GM/c^2$ for the
Schwarzschild metric. Therefore, the differential energy density
per one solid angle of photon direct from accretion disk is
\begin{eqnarray}
u_{\rm disk}(\epsilon,\mu_{\ast};r_{\rm
b})&=&\kappa\frac{3L_{d}}{16\pi^2\eta (R^3/r_{\rm g})}\frac{8\pi
(m_{\rm e} c^2)}{\lambda_{\rm C}^3}\\\nonumber
&&\frac{\epsilon^3}{e^{\epsilon/\epsilon_{\rm
d}(R)}-1}\varphi(R)/(T_{\rm d}^4a)\;, \label{u_disk}
\end{eqnarray}
where $R=r_{\rm b}\sqrt{u_*^{-2}-1}$, the normalization
factor $\kappa=\frac{L_{\rm d}}{\int L_{\rm
bb}(\epsilon)d\epsilon}$, and $T_{\rm
d}=1.164\times10^4\epsilon_{\rm d}(R)m_{\rm e}c^2$ is the
temperature of accretion disk at radius $R$, and $a$ is Boltzmann
energy density constant \citep{dermer09,inoue96}.

For the BLR photon field, its specific energy density depends on
the BLR structure and the radiation feature of accretion disk.
Here, two assumptions are made: (1) the BLR is a spherically
symmetric shell with inner radius $R_i$ and outer radius $R_{\rm
o}$. In general, the radius of BLR can be roughly estimated by
using the relationship between some emission lines luminosity and
the radius of BLR (e.g., Kaspi et al. 2005; Bentz et al. 2009),
and the method of straight-forward reverberation mapping (e.g.,
Paltani \& Turler 2005), but as discussed in \citet{ghisellini08},
these methods also lead to large uncertainties on the radius of
BLR. Fortunately, it has been found that the BLR radius $R_{\rm
BLR}$ can scale roughly as $L_{\rm d}^{0.5}$ (e.g., Wandel et al.
1999) or $L_{\rm d}^{0.7}$ (e.g., Kaspi et al. 2000). Here, we
assume the simplest hypothesis that the BLR radius scales with the
square root of the disk luminosity \citep{ghisellini08}, i.e.,
$R_{\rm BLR} = 10^{17}[L_{\rm d}/10^{45}\ \rm erg\ s^{-1}]^{1/2}\
{\rm cm}$, consider the $R_{\rm BLR}$ as the outer radius $R_{\rm
o}$ of the shell BLR, and assume $\frac{R_{\rm o}}{R_{\rm i}}=\xi$
with $\xi\approx 10$ representing a moderate width; and (2) the
gas density of the BLR has the power-law distribution $n_{\rm
e}(r)=n_0(\frac{r}{R_{\rm i}})^\zeta$ for $ R_{\rm i}\leq r \leq
R_{\rm o}$, it has been suggested that the particle density of the
BLR in quasars scales as $r^{-1.0}$ or $r^{-1.5}$ (e.g., Kaspi \&
Netzer 1999), here we use $\zeta=-1.0$. Since the gas of BLR will
also Thomson scatter the central accretion disk radiation to form
the diffuse radiation from the BLR, the density $n_{\rm e}(r)$ can
be estimated by using the radial Thomson depth $\tau_{\rm
T}=\sigma_{\rm T}\int^{R_{\rm o}}_{R_{\rm i}}dr n_{\rm e}(r)$
\citep{dermer09}, where $\sigma_{\rm T}$ is the Thomson cross
section and $r$ is the distance from the central black hole, here
we use $\tau_{\rm T}=0.01$ which is the typical value (e.g., Donea
\& Protheroe 2003; Finke \& Dermer 2010a). For the radiation field
of standard accretion disk described by the multi-color blackbody
spectrum, its luminosity in the stationary frame is
\begin{equation}
L_{\rm bb}(\epsilon)=8\pi^2 \int^{R_{\rm out}^{\rm d}}_{R_{\rm in}^{\rm d}}\frac{2 c (m_{\rm e} c^2)}{\lambda_{\rm C}^3}\frac{\epsilon^3}{e^{\epsilon/\epsilon_{\rm d}(R)}-1}RdR\,,
\end{equation}
where $\lambda_{\rm C}=h/m_{\rm e}c=2.426\times10^{-10}\ {\rm cm}$
is the electron Compton wavelength and $R_{\rm out}^{\rm
d}=300R_{\rm in}^{\rm d}$ is the outer radius of the accretion
disk. Therefore, the differential energy density per one solid angle of BLR-scattered photon field
is
\begin{equation}
u_{\rm BLR}(\epsilon,\mu_{\ast};r_{\rm b})=\kappa\frac{L_{\rm bb} (\epsilon)r_{\rm
e}^2}{2\pi 3cr_{\rm b}}F(\mu_{\ast},r_{\rm b})\,,
\label{u_BLR}
\end{equation}
where $r_{\rm e}$ is classic electron radius, $r_{\rm b}$ is the
distance from the emission blob to the central black hole, and
$F(\mu_{\ast},r_{\rm b})$ is the function given by
\citet{dermer09} (their Eq.(97)). From equation (\ref{u_BLR}), the
specific energy density of BLR-scattered photon field depends on
the angle, $\theta_{\ast}$, between the directions of the photons
and the electrons, where $\mu_{\ast}= \cos\theta_{\ast}$.

The $\nu \, F_{\rm \nu}$ spectrum of the EC process is given by
\begin{eqnarray}
\nu \, F_{\rm \nu}^{\rm EC(BLR, disk)} &=& 2\pi\frac{c\pi
r_e^2}{4\pi d_L^2}\;\epsilon_s^2\delta_{\rm D}^3\; \int_{-1}^1
d\mu_*\; \\\nonumber
&&\int_0^{\epsilon_{\rm max}} d\epsilon\;
\frac{u_{(\rm BLR, disk)}(\epsilon,\mu_{\ast};r_{\rm b})}
{\epsilon^2}\times
\\\nonumber
& &\int_{\gamma_{\rm low}}^\infty d\gamma\;
\frac{N^\prime_e(\gamma/\delta_{\rm D})}{\gamma^2}\; C_{\rm
kernel}\;, \label{EC_compton}
\end{eqnarray}
where $\epsilon_{\rm s}=h\nu/m_{\rm e}c^2$ and $\gamma=\delta_{\rm
D}\gamma^{\prime}$. The inverse Compton kernel is $C_{\rm kernel}
\;\equiv \; y+y^{-1}  - \frac{2\epsilon_s}{\gamma \bar\epsilon y}
+ (\frac{\epsilon_s}{\gamma \bar\epsilon y})^2$, $y \;\equiv\; 1 -
{\epsilon_s/\gamma}$, and $\bar\epsilon=\gamma\epsilon(1-\mu_*)$
\citep{dermer09}.

For a FSRQ, the thermal emission may be significant, so we also
take the thermal contribution $\nu\, F^{\rm Ther}_{\nu}$ from the
standard accretion disk into account. Therefore, the local
spectrum for total emission from a FSRQ is
\begin{equation}
 \nu \, F^{\rm tot}_{\rm \nu}=\nu\, F^{\rm Ther}_{\nu}+[\nu F^{\rm syn}_{\nu}+\nu F^{\rm
 SSC}_{\nu} + \nu \, F_{\rm \nu}^{\rm EC(BLR, disk)}]\;.
\end{equation}

\subsection{Photon Absorption}

In above descriptions, we did not considered the absorption of
high energy photons. In fact, there are two possible absorption
processes: internal and external. For the internal absorption,
since the photons from BLR and accretion disk not only provide
target radiation fields for IC process to produce $\gamma$-ray
emission, but also attenuate $\gamma$-rays through the
pair-production process, the internal absorption should be taken
into account. The internal absorption optical depth in the
radiation fields of the disk and the BLR.
 is given by
\citet{dermer09}
\begin{eqnarray}
\tau^{\rm int1}(E_{\gamma}, r_{\rm b})&=& 2\pi \int^{\infty}_{\rm
r_{b}}dr\int^1_{-1}(1-\mu_*)d\mu_*\\\nonumber
&&\int_{2/(E_{\gamma}(1+z)(1-\mu_*))}^{\infty} d\epsilon
\sigma_{\gamma\gamma}(s)\times\\\nonumber &&\frac{u_{\rm
disk}(\epsilon,\mu_{\ast};r)+u_{\rm
BLR}(\epsilon,\mu_{\ast};r)}{\epsilon m_{\rm e}c^2} \;,
\label{internal_absorption1}
\end{eqnarray}
where $\sigma_{\gamma\gamma}(s)$ is the $\gamma\gamma$
pair-production cross section (e.g., Gould \& Schr\'{e}der 1967;
Dermer et al. 2009). On the other hand, there is also the
absorption in the blob itself, i.e., the absorption due to the
interaction with the internal synchrotron radiation field (e.g.
Sitarek \& Bednarek 2007). In this case, the internal absorption
optical depth is given by Finke et al. (2008)
\begin{equation}
\tau^{\rm int2}(E_{\gamma})\cong\frac{(1+z)^2\sigma_{\rm T}d^2_{\rm L}}{2m_{\rm e}c^4t_{\rm v,min}\delta_{\rm D}^6}E_{\gamma}\bar{\nu}F_{\bar{\nu}}^{\rm syn}\ ,
\end{equation}
where $\bar{\nu}=\frac{2\delta_{\rm
D}^2}{(1+z)^2E_{\gamma}}\cdot1.236\times10^{20}$ Hz and
$\bar{\nu}F_{\bar{\nu}}^{\rm syn}$ is given by Eq.~\ref{syn}.

\begin{table*}
\caption{List of model parameters used to reproduce the SEDs of 21
FSRQs. The descriptions of columns in Table \ref{tab:tba} are as
follows: column[1]: name; column[2]: redshift; column[3]: magnetic
field in Gauss; column[4--6]: minimum, break and maximum random
Lorentz factors of the relativistic electrons; column[7 and 8]:
slopes of relativistic electrons distribution below and above the
$\gamma^{\prime}_{\rm b}$; column[9]: Doppler factor of the blob;
column[10]: blob size in unit of cm; column[11]: black hole mass
in $10^8$ times solar mass; column[12]: the accretion luminosity;
column[13]: the distance from blob to black hole in cm. $*$: bad
fits at optical or X-ray band, see text for the detailed
comments.} \label{tab:tba}
\begin{center}
\begin{tabular}{lccccccccccccccccccccccccccccccccccc}
 \hline
Name & $z$ & $B$ & $\gamma^{\prime}_{\rm min}$ & $\gamma^{\prime}_{\rm b}$ & $\gamma^{\prime}_{\rm max}$ & $p_1$ & $p_2$ & $\delta_{\rm D}$ & $R^{\prime}_{\rm b}$  & $M_8$ & $\ell_{\rm Edd}$ & $r_{\rm b}$\\
$[1]$ & $[2]$ & $[3]$ & $[4]$ & $[5]$ & $[6]$ & $[7]$ & $[8]$ & $[9]$ & $[10]$ & $[11]$ & $[12]$ & $[13]$\\
 \hline
$0133+47$ & 0.86 & 0.15 & 400 & 3.2E3 & 5.0E4 & 2.0 & 4.0 & 25 & 8.13E16 & 16.1 & 0.03 & 4.07E17\\
$0208-512^{*}$ & 1.0 & 0.18 & 720 & 3.3E3 & 5.0E4 & 2.0 & 3.8 &24 & 1.0E16 & 8.29 & 0.06 & 3.25E17\\
$0227-369^{*}$ & 2.12 & 0.2 & 980 & 3.1E3 & 5.0E4 & 2.0 & 4.5 & 25 & 4.79E15 & 10 & 0.02 & 2.03E17\\
$0347-211$ & 2.94 & 0.22 & 1.86E3 & 3.5E3 &5.0E4 & 2.0 & 4.8 & 23 & 2.1E16 & 10 & 0.06 & 3.99E17\\
$0420-014^{*}$ & 0.92 & 0.18 & 400 & 3.0E3 & 1.5E4 & 2.0 & 5.0 & 26 & 1.23E16 & 32.4 & 0.01 & 3.01E17\\
$0454-234^{*}$ & 1.0 & 0.2 & 810 & 3.1E3 & 4.0E4 & 2.0 & 5.0 & 26 & 4.37E15 & 25.1 & 0.01 & 2.22E17\\
$0528+134$ & 2.07 & 0.24 & 1.3E3 & 2.0E3 & 2.2E4 & 2.0 & 5.0 & 23 & 4.27E15 & 45 & 0.02 & 8.99E17\\
$1454-345$ & 1.42 & 0.24 & 1.1E3 & 2.oE3 & 2.2E4 & 2.0 & 5.0 & 23 & 1.58E16 & 20 & 0.02 & 2.72E17\\
4C 01.28 & 0.89 & 0.24 & 380 & 2.0E3 & 9.0E3 & 2.0 & 4.0 & 23 & 2.0E16 & 17.8 & 0.02 & 2.84E17\\
4C 28.07 & 1.21 & 0.24 & 920 & 2.0E3 & 9.0E3 & 2.0 & 4.0 & 23 & 3.47E15 & 10 & 0.03 & 2.29E17\\
$0917+449^{*}$ & 2.19 & 0.25 & 760 & 2.0E3 & 9.0E3 & 2.13 & 4.0 & 32 & 8.71E15 & 75.8 & 0.05 & 9.81E17\\
4C 29.45 & 0.73 & 0.22 & 760 & 2.0E3 & 3.0E4 & 2.13 & 4.0 & 22 & 4.57E15 & 12.9 & 0.02 & 2.42E17\\
3C 273 & 0.16 & 0.22 & 210 & 800 & 1.25E3 & 2.0 & 5.0 & 16 & 1.7E16 & 10 & 0.4 & 7.6E17\\
$1510-089^{*}$ & 0.36 & 0.22 & 355 & 850 & 1.7E3 & 2.05 & 5.0 & 16 & 3.16E16 & 13.5 & 0.02 & 1.7E17\\
$1308+32$ & 1.0 & 0.18 & 550 & 3.3E3 & 1.2E5 & 2.0 & 3.8 & 24 & 1.38E16 & 3.98 & 0.03 & 1.61E17\\
$1502+106$ & 1.84 & 0.18 & 1.15E3& 3.3E3 & 1.5E5 & 2.0 & 3.8 & 26 & 1.32E16 & 31.6 & 0.02 & 3.13E17\\
$1520+319^{*}$ & 1.49 & 0.22 & 355 & 850 & 3.7E3 & 2.0 & 5.0 & 16 & 3.47E16 & 50 & 0.01 & 1.38E17\\
4C 66.20 & 0.66 & 0.18 & 430 & 3.3E3 & 7.0E4 & 2.0 & 3.8 & 26 & 1.17E16 & 41.5 & 0.01 & 3.09E17\\
OX 169 & 0.21 & 0.18 & 370 & 3.3E3 & 6.3E4 & 2.0 & 3.8 & 26 & 1.48E16 & 9.54 & 0.01 & 2.2E17\\
$2325+093$ & 1.84 & 0.15 & 1.4E3 & 2.0E3 & 2.2E4 & 2.0 & 5.0 & 23 & 5.37E15 & 20 & 0.03 & 3.3E17\\
$2345-1555^{*}$ & 0.62 & 0.15 & 1.23E3 & 2.0E3 & 2.2E4 & 2.0 & 4.0 & 23 & 2.14E15 & 10 & 0.01 & 1.52E17\\
  \hline
\end{tabular}
\end{center}
\end{table*}

For the external absorption, because high energy photons emitted
from any FSRQ will interact with extragalactic background light
(EBL) in the propagation in the extragalactic space and then are
absorbed through the pair-production process (e.g., Stecker et al.
1992; Finke et al. 2010b), such an absorption should be taken into
account. Here we adopt the EBL model of \citet{finke10} to get the
optical depth of a $\gamma$-ray photon with energy $E_{\gamma}$ at
redshift $z$, $\tau^{\rm EBL}(E_\gamma,z)$.

Therefore the photons flux observed at the Earth is
\begin{eqnarray}
\nu\,F_{\nu} &=& (\nu\,F^{\rm tot}_{\nu})\exp[-\tau^{\rm
EBL}(E_{\gamma},z)-\tau^{\rm int1}(E_{\gamma}, r_{\rm
b})]\\\nonumber &&\times\left(\frac{1-\exp(-\tau_{\rm
int2})}{\tau_{\rm int2}}\right)\;,
\end{eqnarray}
where $\nu\,F^{\rm tot}_{\nu}$ is given by Eq. (9).

\section{Applications}

\begin{figure*}
\begin{center}
\includegraphics[width=140mm]{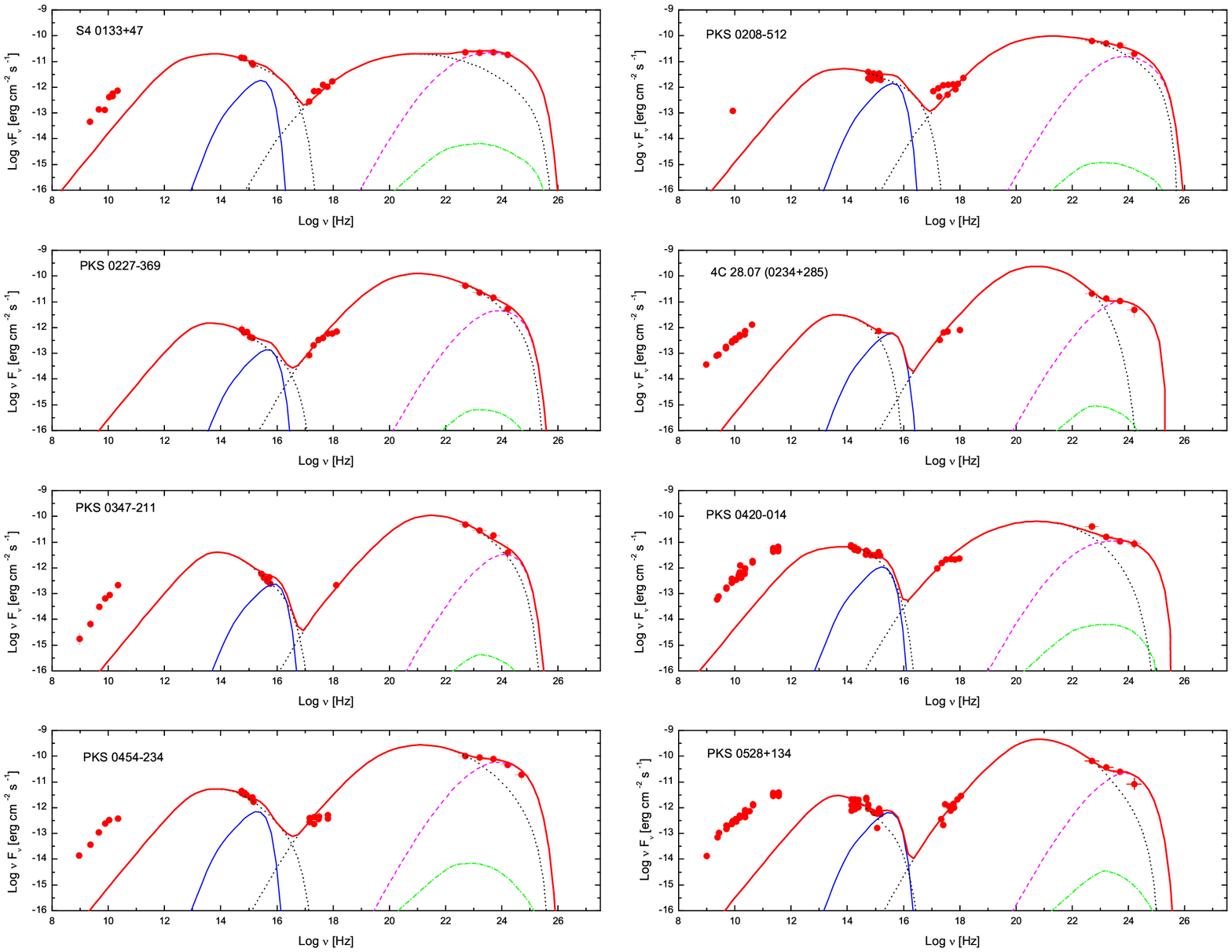}
\end{center}
\caption{Comparisons of modelling SEDs with the observed data for
the Fermi FSRQs.  For each FSRQ, thin solid line represents the
thermal emission from accretion disk, dotted, dashed, and
dash-dotted lines represent SSC, BLR-Compton, and  disk-Compton
components, respectively, and bold solid line is the sum of these
components. The observed data are taken from \citet{abdo10a}.}
   \label{Fig1}
\end{figure*}

\begin{figure*}
\begin{center}
\includegraphics[width=140mm]{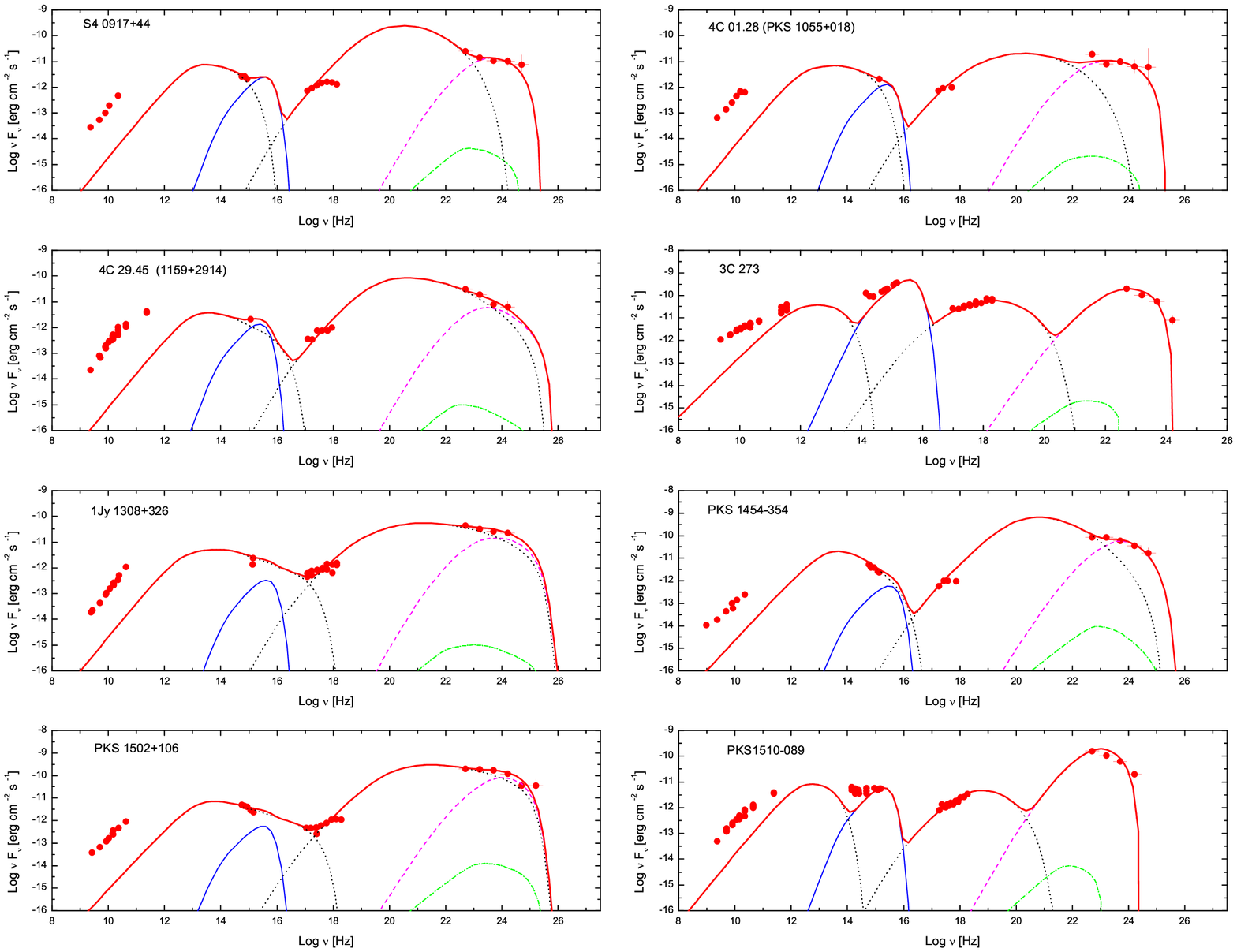}
\end{center}
\caption{Same as Fig.~\ref{Fig1}, but for different FSRQs. }
   \label{Fig2}
\end{figure*}

\begin{figure*}
\begin{center}
\includegraphics[width=140mm]{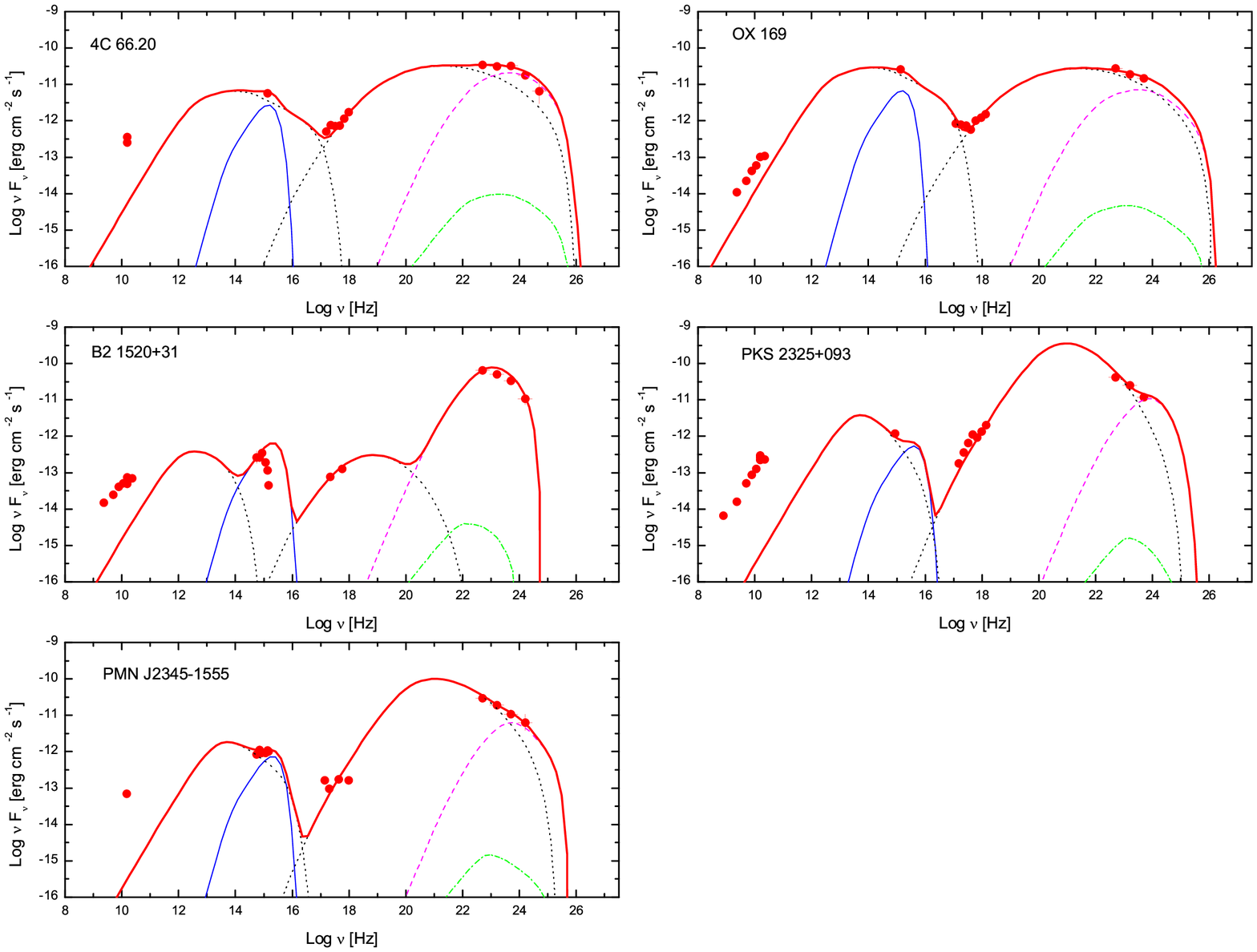}
\end{center}
\caption{Same as Fig.~\ref{Fig1}, but for different FSRQs.}
   \label{Fig3}
\end{figure*}

We now apply the model describes in Section 2 to reproduce the
quasi-simultaneous multi-frequency SEDs of 21 FSRQs. The observed
data of these 21 FSRQs are taken from \citet{abdo10a}. In fact,
\citet{Giommi11} construct simultaneous multi-frequency SED of 105
blazars, however, most of these blazars were observed for short
time, and the numbers of collected high energy photons are small.
Hence, most of these blazars only have upper limit of the GeV
flux, which are not helpful for our goal. Six FSRQs in the sample
of \citet{Giommi11} having three or four GeV data points are
0420-01, 0454-234, 4C 29.45, 3C 273, 1308+326 and 1502+106, which
are also presented in the sample of \citet{abdo10a}.

The modelling results are shown in Figs.~\ref{Fig1}--~\ref{Fig3}.
It can be found that our models can reproduce well the
quasi-simultaneous emission from optical--X-ray to GeV band of
most FSRQs. Our model can not reproduce very well the optical or
X-ray spectrum of several sources (marked with $*$ in Table
\ref{tab:tba}). We believe that the bad reproduction at optical
band of PKS 1510-089 is caused by the fact that the observed
optical data are contaminated by the emission from its host
galaxy. For the source B2 1520+31, it seems that the standard
accretion disk model we used here does not adapt to describe its
UV emission. The observed X-ray spectrum of these sources
0208-512, 0227-369, 0420-014, 0454-234, 0917+449, and 2345-1555
become flat at several keV band, which can not be reproduced by
our model. In fact, the relativistic electron distribution with a
large $\gamma'_{\rm min}$ can possibly explain such a spectrum
\citep{tavecc09}. The model parameters are listed in Table
\ref{tab:tba}. The same values of $\eta=1/12$, $\zeta=-1$,
$\tau_{\rm T}=0.01$ and $\frac{R_{\rm o}}{R_{i}}=10$ are used for
all sources in our sample, which are not listed in Table
\ref{tab:tba}. The black hole masses of 0420-014, 1510-089 and 4C
66.20 (1849+64) are taken from \citet{gu01}. The black hole masses
of 4C 01.28 (1055+018), 0917+449, 4C 29.45, 1308+32, 1502-106 and
OX 169 are taken from \citet{chen09}. The black hole masses of the
other sources are derived through modelling SEDs. It should be
noted that the thermal emission of PKS 1510-089, 3C 273 and B2
1520+31 are so significant that the optical-UV emission are purely
from the thermal emission. Meanwhile, they have the smallest value
of $\gamma^{\prime}_{\rm b}$ and $\gamma^{\prime}_{\rm max}$. This
indicates that the environment around the blob have a significant
effect on the distribution of electrons in blob and the cooling of
the electrons are dominated by EC process for these sources.

From Table 1, we can find that the black hole masses are in the
range (1--6)$\times10^9M_{\rm sun}$, the accretion luminosity of
most of our sources are in the range 0.01 -- 0.05 (only one source 3C
273 requires the accretion luminosity larger than 0.1), and the
disk luminosity cluster around $6\times10^{45} \rm \ erg\ s^{-1}$.
For the high energy electron's distribution, we find that the
values of the $\gamma^{\prime}_{\rm min}$ are in the range
400--1500, which are quite larger than the results derived by
\citet{ghisellini10}, indicating some clues to the acceleration
and cooling processes of the electrons \citep{tavecc09,kata}. From
Table 1, it's found that $\gamma^{\prime}_{\rm max}$ clusters
around $3.0\times10^{4}$, and besides three sources with
$\gamma^{\prime}_{\rm b}\sim800$, 8 sources have
$\gamma^{\prime}_{\rm b}\sim2000$ and 10 sources have
$\gamma^{\prime}_{\rm b}\sim3000$. For the magnetic field of
radiation region, their values are distributed in a narrow range
0.15 -- 0.25 G, which are far smaller than the results found by
\citet{ghisellini10}. The small magnetic field means inefficient
synchrotron cooling, which are consistent with the results of
large $\gamma^{\prime}_{\rm min}$ and $\gamma^{\prime}_{\rm b}$.
The values of Doppler factor $\delta_{\rm D}$ we derived range
from 16 to 32, which is in the range reported by
\citet{Savolainen10}.

\begin{figure}
\begin{center}
\includegraphics[width=90mm]{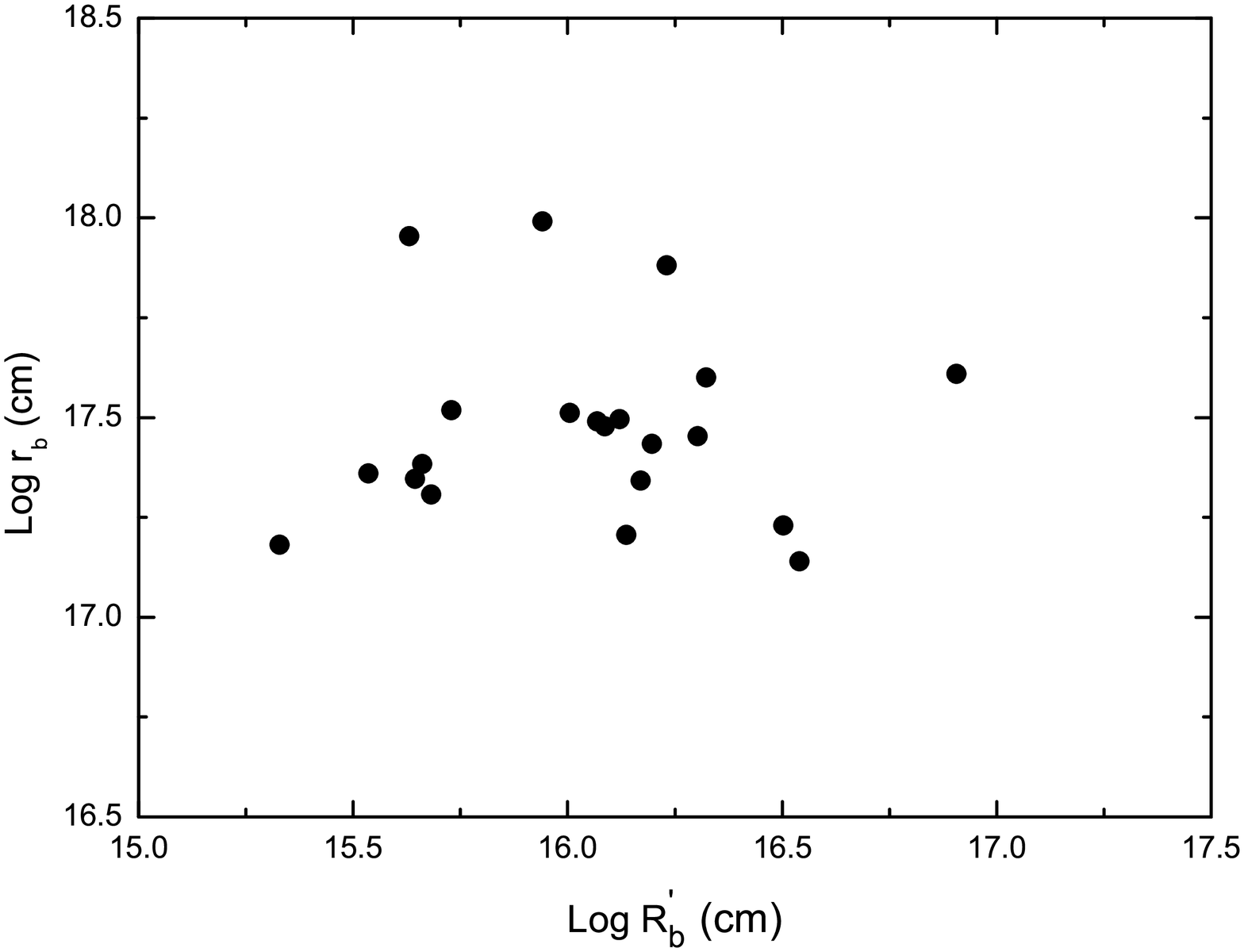}
\end{center}
\caption{The relationship between the location of the blob and the
size of the blob. }
   \label{location-size.eps}
\end{figure}

\begin{figure}
\begin{center}
\includegraphics[width=90mm]{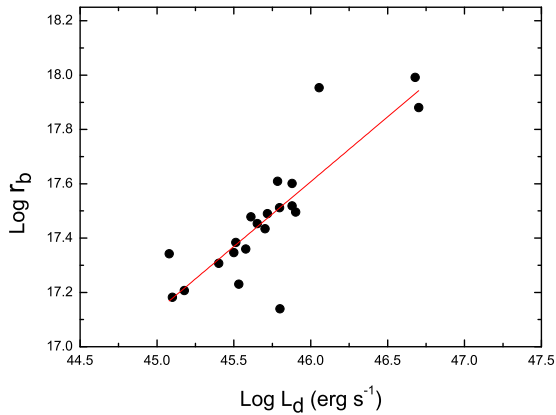}
\end{center}
\caption{The relationship between the location of the blob and the
accretion disk luminosity. }
   \label{location-disk}
\end{figure}

\begin{figure}
\begin{center}
\includegraphics[width=90mm]{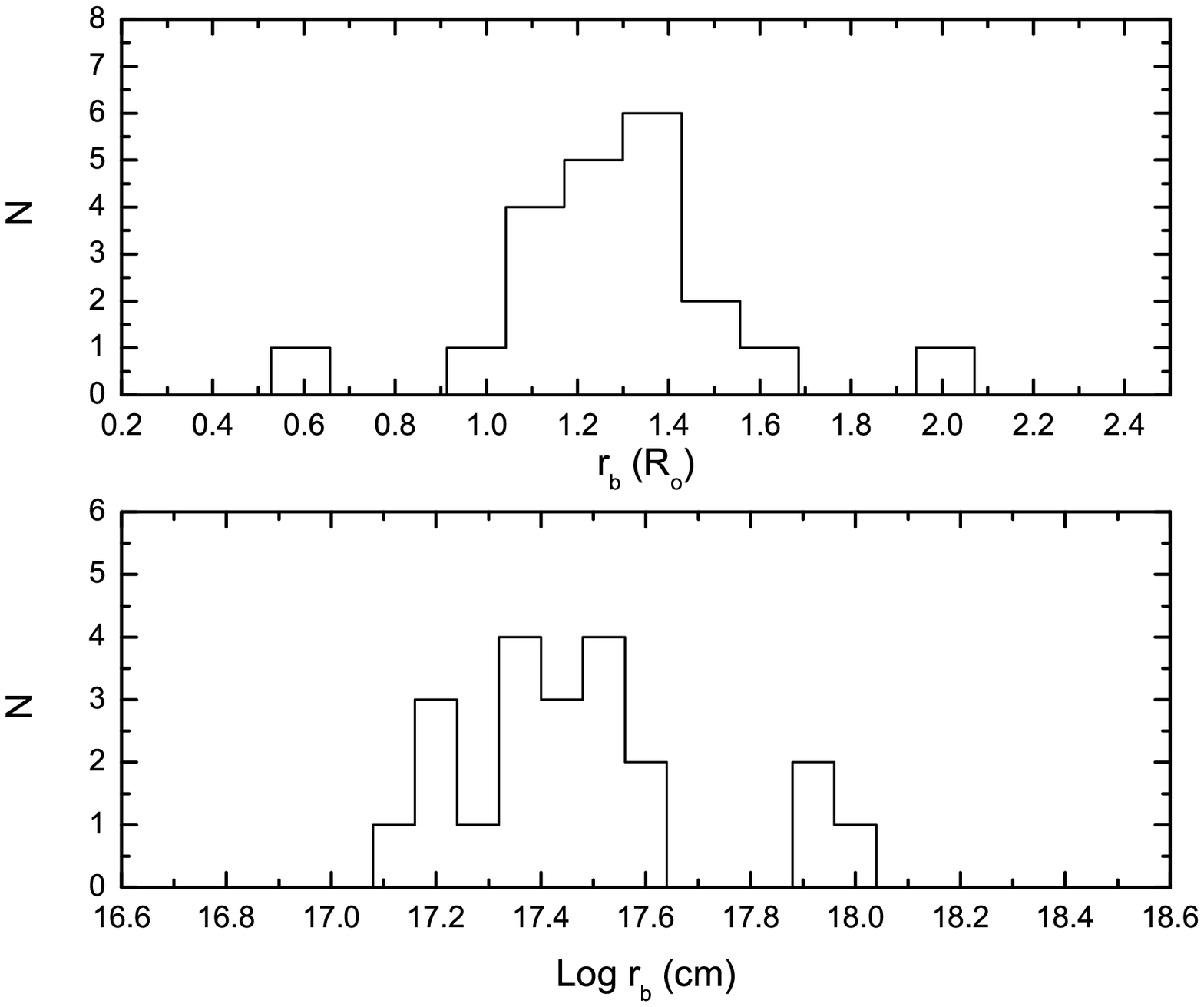}
\end{center}
\caption{Distributions of the locations of emission regions. Top
panel: $r_{\rm b}$ is in unit of outer radius of BLR $R_{\rm o}$.
Bottom panel: $r_{\rm b}$ is in unit of cm. }
   \label{Fig6}
\end{figure}

We now consider the relationships among size $R^{\prime}_{\rm b}$
of the blob, $L_{\rm d}$, and the location $r_{\rm b}$ of the
blob. It is expected that $r_{\rm b}\theta_{\rm j}$ relates to
$R^{\prime}_{\rm b}$ by $R^{\prime}_{\rm b}=r_{\rm b}\theta_{\rm
j}$, where $\theta_{\rm j}$ is the opening angle of jet. However,
we find that there is no correlation between $R^{\prime}_{\rm b}$
and $r_{\rm b}$ (see Fig.~\ref{location-size.eps}). The possible
large distribution of $\theta_{\rm j}$ in our sample would disturb
the expected correlation. On the other hand, this fact may support
either the scenario of the $\gamma$-ray produced in a more compact
region embedded in jet \citep{Giannios09, tavecc11} or the
re-collimation of the flow \citep{brom09}. In our sample, we find
that there is a strong correlation between $r_{\rm b}$ and $L_{\rm
d}$ with correlation coefficient $r=0.83$ and a chance probability
$p=1.94\times10^{-6}$ (Fig.~\ref{location-disk}). This correlation
is caused by the fact that $r_{\rm b}$ cluster around $1.3R_{\rm
o}$ (top panel in Fig.~\ref{Fig6}).

Our results confirm the well known cognitive that EC process is
important for $\gamma$-ray emissions of the FSRQs. The
contribution of EC component to $\gamma$-ray emission of the FSRQ
is dominated by BLR component, which is determined by the location
of the emission region (blob) $r_{\rm b}$ and BLR structure.
Therefore, $\gamma$-ray emission of the FSRQ can be used to
constrain the location of emission blob. In Fig.~\ref{Fig6}, we
present the distribution of the locations of emission regions
$r_{\rm b}$ of 21 FSRQs. We find that all these sources have
$r_{\rm b}>0.5R_{\rm o}$ and the emission region of 19 out of 21
are outside of the BLR with the typical assumptions of BLR
structure, i.e., their blobs are located in the distance range of
$10^{17}\sim10^{18}\ {\rm cm}$.

\begin{figure}
\begin{center}
\includegraphics[width=90mm]{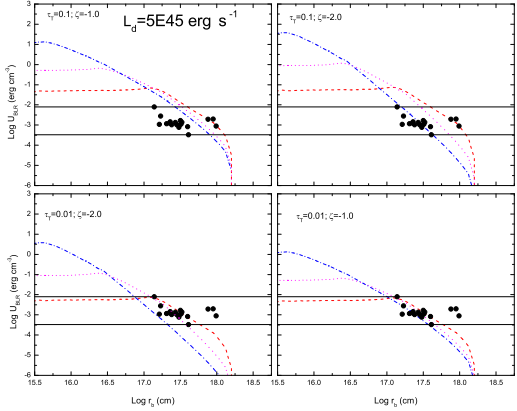}
\end{center}
\caption{Change of energy density of BLR-scattered photon field
with the location $r_{\rm b}$ of emission region for different
assumptions of the BLR structure. The black points are the value
obtained from the SED fits of modelled sources. The horizonal
lines are the lower and upper limits of $U_{\rm BLR}$ derived from
the $\gamma$-ray emission of our sample, respectively. Dashed,
dotted, and dash-dotted lines correspond to the cases of $R_{\rm
o}/R_{\rm i}=2$ , 10, and 60, respectively. For accretion
luminosity and black hole mass, the typical values found in our
sample are used: $M_{\rm BH}=2.0\times10^9\ M_{\rm sun}$ and
$\ell=0.02$. }
   \label{Fig7}
\end{figure}

\begin{figure}
\begin{center}
\includegraphics[width=90mm]{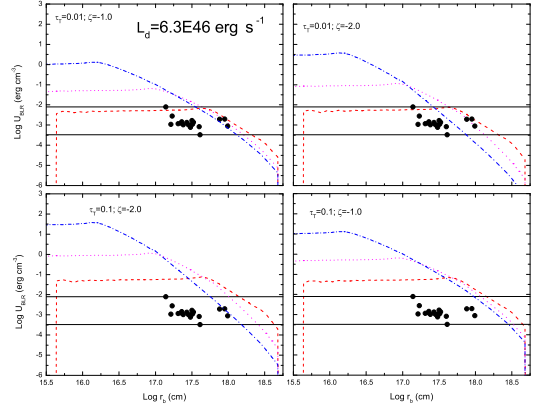}
\end{center}
\caption{Change of energy density of BLR-scattered photon field
with the location $r_{\rm b}$ of emission region for different
assumptions of the BLR structure. The symbols are the same as that
in Fig.~\ref{Fig7}. For accretion luminosity and black hole mass,
$M_{\rm BH}=5.0\times10^9\ M_{\rm sun}$ and $\ell=0.1$ are used. }
   \label{Fig8}
\end{figure}

Since the EC component relates to the differential energy density per one solid angle $u_{\rm BLR}$
of BLR photon field and $u_{\rm BLR}$ depends on the BLR structure.
We now discuss the uncertainty due to the parameters ($\zeta$,
$\tau_{\rm T}$, and $R_{\rm o}/R_{\rm i}$) describing the BLR
structure. For these three parameters, there is little observed
information. Except for the typical values of BLR structure we used,
there are alternative values for these parameters. In a wind model
for BLR suggested by \citet{murray} and \citet{elvis}, the value of
$\zeta$ should be -2. The alternative extreme values of $\tau_{\rm
T}$ is 0.1 \citep{dermer09}. The width of the BLR is poorly known,
we will consider the cases of $R_{\rm o}/R_{\rm i} = (2,10,60)$,
which are corresponding to a very thin BLR shell, a moderate thin
BLR shell and a thick BLR shell, respectively. From the observed
$\gamma$-ray flux, the energy density of BLR photon field required
to reproduce the SEDs can be derived,
\begin{equation}
U_{\rm BLR}(r_{\rm b})=\int d\mu_* \int u_{\rm BLR} (\epsilon,
\mu_*; r_{\rm b})d\epsilon\;.
\end{equation}
We find that the values of $U_{\rm BLR}(r_{\rm b})$ are restricted
in the range of $(3.2\times10^{-4}$\ --\ $7.9\times10^{-3})\, \rm
erg\ cm^{-3}$ in our sample. Hence, no matter what assumption
about the BLR structure we make, the allowed location of emission
region would make the value of $U_{\rm BLR}(r_{\rm b})$ in the
above range. In Fig.~\ref{Fig7}, we show the relationship between
energy densities of BLR-scattered photon field with different
parameters for the BLR structure and the locations of emission
regions with typical values in our sample, where we assume that
$M_{\rm BH}=2.0\times10^9\ M_{\rm sun}$ and $\ell=0.02$ (i.e.
$L_{\rm d}\approx5\times10^{45}\ \rm erg\ s^{-1}$). We find that
the smallest allowed value of $r_{\rm b}\approx7.9\times10^{16}$
cm is derived when $\tau_{\rm T}=0.01$, $\zeta=-2$ and $R_{\rm
o}/R_{\rm i}=60$. With $\tau_{\rm T}=0.1$, $\zeta=(-1, -2)$ and
$R_{\rm o}/R_{\rm i}=2$, the largest allowed value of $r_{\rm
b}\approx1.3\times10^{18}$ cm is obtained. Since there are three
sources with high accretion disk luminosity ($L_{\rm d}>10^{46}
\rm \ erg \ s^{-1}$) in our sample, and the accretion disk
luminosity has a affect on constraining the location of blob, we
also discuss the case of high accretion disk luminosity $L_{\rm
d}\approx6.3\times10^{46}\ \rm erg\ s^{-1}$ with $M_{\rm
BH}=5.0\times10^9\ M_{\rm sun}$ and $\ell=0.1$. The results are
shown in Fig.~\ref{Fig8}. It is found that the allowed value
of $r_{\rm b}$ is in the range $2.6\times10^{17}$\ --\
$4.2\times10^{18}$\ cm.

\begin{figure}
\begin{center}
\includegraphics[width=90mm]{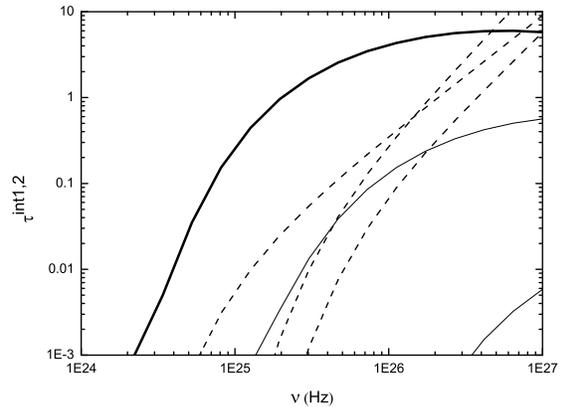}
\end{center}
\caption{The optical depths of internal absorption in modeling the
SEDs of FSRQs. Solid line: the absorption in the accretion disk
and BLR radiation fields $\tau^{\rm int1}$ (from left to right: B2
1502+106, PKS 0528+134, and OX 169, which has the largest,
intermediate and the lowest optical depth of $\tau^{\rm int1}$,
respectively. B2 1502+106, the source with the strongest
absorption in the accretion disk and BLR radiation fields, is
marked with a thick curve.); Dashed line: the absorption in blob
$\tau^{\rm int2}$ (from left to right: PKS 0528+134, PKS 0347-211,
and PKS 0227-369, which have the largest optical depths of
$\tau^{\rm int2}$ ). See discussion in text.}
   \label{Fig9}
\end{figure}

Finally, we consider the roles of both internal and external
absorptions. For the internal absorptions in our sample, we find
that they are negligible in reproducing the SEDs. As examples, in
Fig.~\ref{Fig9}, we show the internal optical depths in the
radiation fields of the disk and the BLR for three sources (B2
1520+31, PKS 0528+134, and OX 169) and those in the synchrotron
radiation fields for PKS 0528+134, PKS 0347-211, and PKS 0227-369.
From Fig.~\ref{Fig9}, B2 1520+31 has a largest value of $\tau^{\rm
int1}$ due to its high energy density of external photon
field, $U^{\rm max}_{\rm BLR}(r_{\rm b}^{\rm
min})=7.9\times10^{-3}\ \rm erg\ cm^{-3}$, which becomes
significant for photons with $\nu>10^{25}$\ Hz. As to PKS 0528+134
and OX 169 in Fig. \ref{Fig9}, the former has the intermediate
optical depth and the latter has the lowest optical depth in our
sample. We expect that the $\gamma$-ray spectrum between 0.1\ --\
10\ GeV could not be reproduced well if the energy density of BLR
photon field exceeds $U^{\rm max}_{\rm BLR}(r_{\rm b}^{\rm min})$.
For the absorption in blob itself, we show largest values of the
optical depths, $\tau^{\rm int2}$, in our sample, it can be seen
that these absorptions are negligible in our sample. On the other
hand, the fact that there is no significant spectral cutoff
between 10\ -\ 100\ GeV for FSRQs
\citep{Costamante2011,Ackerman11} also supports the minimum value
of $r_{\rm b}$ we derived. For the external absorption, it depends
on the EBL models, so different models will make some differences,
in particular, for higher redshifts. We found that 15 sources have
the maximum gamma-ray energy $E_{\rm max}\sim 6.6$ GeV, 5 sources
have $E_{\rm max} \sim 20.8$ GeV, and 1 source has $E_{\rm
max}\sim 65.7$ GeV in our sample. Therefore, we calculate the
external optical depths for six sources (4C 66.20, 4C 01.28, PKS
0454-234, PKS 1454-354, PKS 1502+106, and S4 0917+44 ) with
$E_{\rm max} > 20$ GeV using four kinds of the EBL models given by
Kneiske et al. (2004), Stecker et al. (2006), Franceschini et al.
(2008), and Finke et al. (2010), respectively, and show the
results in Fig. 10. It can be seen that our modelled SEDs for
three sources with $z\le 1$ are not affected by the EBL models,
but the SEDs for the sources (PKS 1454-354, PKS 1502+106, and S4
0917+44 ) with $z>1.4$ will be different for different EBL models,
particularly, for the EBL model given by Stecker et al. (2006).
However, the EBL model given by Stecker et al. (2006) has been
ruled out by Fermi-LAT observations (Abdo et al. 2010c). For PKS
1502+106, the optical depth given by Kneiske et al. (2004) could
be slightly larger than that expected by Finke model, which would
lead to slightly difference for modeling the SED of PKS 1502+106.
Considering the specific SED of PKS 1502+106, we expect that this
difference has no effect on our results.

\section{Conclusions}

In this paper, we have modeled the quasi-simultaneous multi-band
spectra of 21 {\it Fermi} FSRQs in the frame of a multi-component
one-zone leptonic emission model, and studied the locations of
emission regions of the FSRQs in particular. We have found that
the emission regions lie in the region of $7.9\times10^{16}$\ --\
$1.3\times10^{18}$\ cm (300 -- 4300 Schwarzschild radii) for FSRQs
with low accretion disk luminosity, and the emission regions
locate in the larger region of $2.6\times10^{17}$\ --\
$4.2\times10^{18}$\ cm (300 -- 5600 Schwarzschild radii) for ones
with high accretion disk luminosity. Our results are consistent
with the results of \citet{tavecc10} derived by analyzing the
variability timescale and that of \citet{Liu08} and \citet{bai}
obtained by discussing internal absorption. Our results disfavor
the far dissipation scenario suggested by e.g.,
\citet{Marscher10}, in which $\gamma$-rays are produced at larger
distance ($>$10\ pc) from black hole. At such large distance, the
external radiation field for IC process are dominated by the
emission from the dusty torus with $\nu_{\rm
IR}=3.0\times10^{13}$\ Hz and $U_{\rm dusty}\leq3\times10^{-4} \rm
\ erg\ cm^{-3}$ \citep{ghisellini08}. As we discussed above, the
observed $\gamma$-ray flux requires that the energy density of
external radiation field for IC process should be larger than
$3.2\times10^{-4} \rm \ erg\ cm^{-3}$ with the moderate Doppler
factor, which is inconsistent with the far dissipation scenario.
Moreover, the observed variability on timescale of\ $\sim$\ 10\
min of PKS 1222+216 \citep{Aleksi11,tavecc11} requires quite fast
IC cooling, which also challenges the far dissipation scenario
considering lower energy infrared photons as seed photons.

\begin{figure}
\begin{center}
\includegraphics[width=90mm]{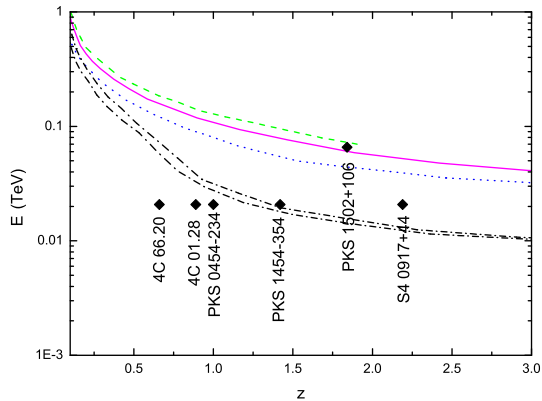}
\end{center}
\caption{Change of the photon energy at $\tau^{\rm EBL}(E_\gamma,
z)=1$ with redshift $z$ for different EBL models. Solid, dashed,
dotted, and dot-dashed lines represent the results by using the
models given by Finke et al. (2010), Franceschini et al. (2008),
Kneiske et al. (2004), and Stecker et al. (2006). Filled diamonds
represent the maximum $\gamma$-ray photon energy for six sources
(4C 66.20, 4C 01.28, PKS 0454-234, PKS 1454-354, PKS 1502+106, and
S4 0917+44 ).}
   \label{Fig10}
\end{figure}

We did not consider the target photons from BLR emission line in
this work. Luminosity of the BLR emission line is comparable with
or larger than that of the BLR Thomson scatted photons radiation
between $10^{15}$\--$\ 10^{16}$\ Hz in the BLR zone
\citep{tavecc08,Sbarrato11}, which would make the locations of the
emission regions we derived closer to the central black hole than
the real ones. Based on the above discussion, we can roughly
estimate that how much the emission line radiation affect our
results. From our results, we can see that when the energy density
of BLR-scattered photons field increases by $\sim$10 times
(corresponding to $L_{\rm d}$ varying from $5\times10^{45}$ to
$6.3\times10^{46}$ $\rm erg\ s^{-1}$, Figs. \ref{Fig7} and
\ref{Fig8}), the range of the emission region locations increases
by $\sim$3 times. Therefore, if the energy density of the BLR
emission line photon field is $\sim$10 times that of the BLR
Thomson scatted photons field, we expect the locations of the
emission regions possibly will locate in the range
$3.0\cdot(7.9\times10^{16}$\ --\ $1.3\times10^{18})$\ cm and
$3.0\cdot(2.6\times10^{17}$\ --\ $4.2\times10^{18})$\ cm for low
and high accretion luminosity FSRQs, respectively. The detail
contribution of BLR emission line will be studied in the future
work.

\bigskip

\section*{Acknowledgments}
This work is partially supported by  a 973 Program (2009CB824800),
the National Natural Science Foundation of China (NSFC 10963004,
11063003), and Yunnan Province under a grant 2009 OC.


\end{document}